# Two-Dimensional Conformal Plasma Turbulence in the Hasegawa-Mima Equation

Shigeo Kawata[1, 2]

[1]Utsunomiya University, Graduate School of Regional Development and Creativity
Yohtoh 7-1-2, Utsunomiya 321-8585, Japan
e-mail address: kwt@cc.utsunomiya-u.ac.jp
[2] Shanghai Jiao Tong University, Laboratory for Laser Plasmas,
800 Dong Chuan Road, Shanghai 200240, China

The two-dimensional (2D) conformal field theory (CFT) suggests that the 2D plasma turbulence, governed by the Hasegawa-Mima (H-M) equation, may have multiple exponents of energy spectrum in momentum space. Electrostatic potential driven by drift waves in magnetized 2D plasmas would be described by the H-M equation. On the other hand, the 2D CFT has an infinite-dimensional symmetry. When we focus on minimal models established in 2D CFT, each minimal model provides a different 2D statistical model as presented in fluid turbulence, quantum field theory and string theory, and would provide a specific exponent of the energy spectrum. The CFT analytical results in this work suggests that the H-M plasma turbulence may have multiple exponents of the energy spectrum.

In plasma and fluid, turbulence is abundant and has been studied extensively [1, 2]. From perturbations, complex flows and waves would appear with vortices in plasma and fluid. Through instabilities, plasmas and fluid may move to turbulent state. For the high Reynolds number in fluid, Richardson introduced the energy cascade, in which energy is injected into "big whirls" [3]. The energy is transferred to "little whirls" and is dissipated by viscosity. Kolmogorov proposed the energy spectrum for spatially homogeneous and isotropic turbulence by a dimension analysis for the energy transport range, that is, called the inertial range [4, 5]: $E(k) \propto \varepsilon^{2/3} k^{-5/3}$. Here $\varepsilon$ the energy flux and $k$ the wavenumber. In addition, for many years two-dimensional (2D) fluid turbulence has attracted scientists [6], and it is known that 2D inviscid fluid equation has an infinite number of the Casimir invariants [7]. It was suggested in 2D fluid turbulence that the energy is transferred to the inverse direction from small scales to large scales, though the enstrophy is transferred from large scales to small scales [8-10].

In plasmas, 2D plasma turbulence has been also focused relating to plasma confinement, structure formation and science interest. For example, the Hasegawa-Mima (H-M) equation was derived for electrostatic potential $\varphi(x,y)$ driven by drift waves in magnetized 2D plasmas [11]:

$$\partial_t(\nabla^2 \varphi - \varphi) - \nabla\varphi \times \hat{z} \cdot \nabla\left(\nabla^2\varphi + \ln\frac{\omega_{ci}}{n_0}\right) = 0 \quad (1)$$

Here $\partial_t = \frac{\partial}{\partial t}$, $\nabla = (\partial_x, \partial_y, 0)$, the time $t$ is normalized by $\omega_{ci}^{-1}$, $\omega_{ci}$ the ion cyclotron frequency, the space is normalized by the ion cyclotron radius $\rho_s$ and $n_0$ shows the plasma unperturbed ion number density. In this case, the plasma is magnetized by a strong magnetic field in the $z$ direction. Figure 1 shows an example physical setup in the H-M equation. In Fig. 1 a uniform density gradient is in the $y$ direction, the static magnetic field is in $z$ and a cyclotron motion is shown for several ions. By the density gradient in $y$, a diamagnetic current appears [12]. When transverse perturbations of the static electric field $(E_x, E_y)$ and/or of the density appear, additional diamagnetic current is induced. The $\vec{E} \times \vec{B}$ drift and the polarization drift in the $x-y$ plane contribute to drift instability, which would move to a turbulence.

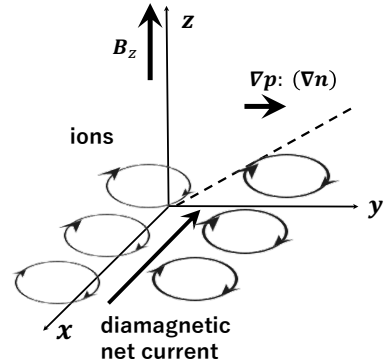

Fig. 1. An example setup for plasma governed by the Hasegawa-Mima equation.

On the other hand, the 2D conformal field theory (CFT) has an infinite conformal symmetry [13]. When we focus on minimal models established in 2D CFT, each minimal model provides a different 2D statistical model as presented in fluid turbulence, quantum field theory and string theory, and would provide a specific exponent of the energy spectrum [13-15]. When a steady state in plasma turbulence is focused in the H-M equation, 2D CFT works and would suggest that 2D plasma turbulence governed by the H-M equation may have multiple exponents of energy spectrum.

In this work a boundary CFT is employed to study the 2D plasma turbulence by the H-M equation [16-18]. The analytical result in the paper presents multiple exponents of the energy spectrum in the 2D plasma turbulence by the H-M equation.

When we focus on a steady state in the H-M equation above,

$$\nabla^2 \varphi + \ln \frac{\omega_{ci}}{n_0} = constant, \qquad (2)$$

and the Green function provides a solution for $\varphi$: $\varphi \propto \ln r$, where $r = \sqrt{x^2 + y^2}$. Therefore, $\varphi$ itself is not a primary field, but $\psi_\mu \equiv \partial_\mu \varphi$, ($\mu = x$ or $y$) would be primary fields in CFT [13]. The H-M equation also shows the following invariants of energy $E$ and enstrophy $\Omega$ [2, 11]:

$$E = \iint d^2x \, \{\varphi^2 + (\nabla \varphi)^2\} \qquad (3)$$

$$\Omega = \iint d^2x \, \{(\nabla \varphi)^2 + (\nabla^2 \varphi)^2\} \qquad (4)$$

In the momentum space, the energy density may become $E_k \sim (1/k^2 + 1)\psi^2$, where $\psi^2 = \psi_\mu \psi^\mu$.

In 2D CFT, Polyakov has developed the minimal model, which has a finite number of operators [13]. The minimal models were also applied to describe 2D inviscid fluid turbulence with the simple Hopf equation with a constant enstrophy flux [14, 15]. The boundary CFT (BCFT) was then developed to include the boundary effect in Refs. [16, 17], and was also applied to fluid turbulence [18]. Now a completely new approach was introduced to plasma turbulence by 2D CFT and BCFT. As shown in Fig. 2, in BCFT mirror fields are introduced in $y < 0$ in this work, and the boundary is located at $y = 0$.

In Ref. [13], the minimal model $\mathcal{M}_{p,q}$ shows the conformal dimension $\Delta_{p,q}$ for the primary operators of $\psi_{\mu; m,n}(z), (z = x + iy)$ at present in the H-M equation:

$$\Delta_{r,s} = \frac{(ps - qr)^2 - (p - q)^2}{4pq} \qquad (5)$$

Here $1 \leq r < p$ and $1 \leq s < q$. In BCFT, a boundary effect is included in CFT by images [16-18]. For example, at $y = 0$ a boundary is introduced as shown in Fig. 2, and in the negative $y$ image fields are imposed so that no transverse electric field appears across the boundary. In addition, an $i$-point correlation function in the domain $y > 0$ is obtained by a $2i$-point correlation function. In order to obtain $E_k$ and $\Omega_k$ in the H-M plasma turbulence, the two-point function of $<\psi^2> = <\psi_\mu(z_1)\psi_\mu(z_2)>$ should be obtained, and the operator product expansion (OPE) is used [13] in 2D BCFT:

$$<\psi_\mu(z_1)\psi_\mu(z_2)> = $$
$$\sum_j C^{\phi_j}_{\psi\psi} \delta z^{2(\Delta_{\phi_{\mu;j}} - 2\Delta_{\psi_\mu})} <\phi_{\mu;j}> + secondaries \qquad (6)$$

Here $\phi_{\mu;j}$ shows a primary field with its dimension of $\Delta_{\phi_{\mu;j}}$ from the OPE of $[\psi_\mu][\psi_\mu]$, the "secondaries" mean the descendant fields of the conformal family $[\phi_{\mu;j}]$ in 2D CFT, and $\delta z = |z_1 - z_2|$. When $\delta z$ become small, the two-point correction should damp. Therefore,

$$\Delta_{\phi_{\mu;j}} > 2\Delta_{\psi_\mu}. \qquad (7)$$

As found in Refs. [15, 18], the one-point function is not always zero, and this is different from the unitary CFT. The one-point function of $<\phi_{\mu;j}>$ is obtained inversely by the two-point function in BCFT as follows:

$$<\phi_{\mu;j}> \propto \delta z^{-2\Delta_{\phi_{\mu;j}}} \qquad (8)$$

When $y \gg 0$, $\delta z = |z_1 - z_2| \sim 2y$. In the 2D plasma turbulence by the H-M equation, we expect to have large scale structures, and in this case the one-point function would not be zero.

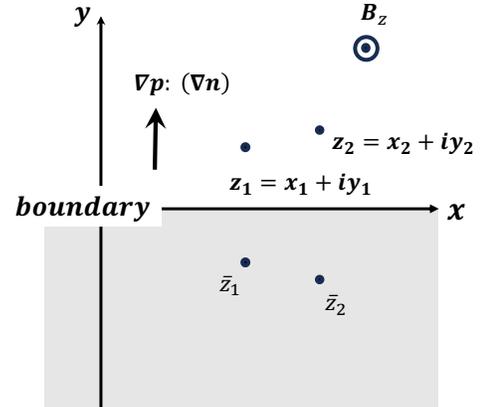

Fig. 2. The Boundary CFT employed in this work. The boundary is located at $y = 0$. In $y < 0$, mirror fields are considered.

The energy spectrum is now obtained below. When we focus on a region far from the boundary ($y \gg 0$), the first term in Eq. (6) is dominant and it is physically reasonable to assume that the exponents of $\Delta_{\psi_\mu}$ and $\Delta_{\phi_{\mu;j}}$ do not depend on the spatial direction of $\mu$. The strong magnetic field in z would contribute to rotate the turbulence so that the isotropic assumption would be reasonable in the H-M plasma turbulence. In this case, the plasma turbulence becomes isotropic. Then, the energy spectrum is estimated as follows:

$$E_k \sim (1/k^2 + 1)\psi^2{}_k \propto (1/k^2 + 1)k^{4\Delta_\psi - 2\Delta_\phi + 1}$$
$$\propto (1 + k^2)k^{4\Delta_\psi - 2\Delta_\phi - 1} \quad (9)$$

Here $\psi^2{}_k$ shows the Fourier transform of $\psi^2$. When $k > 1$, that means the wavelength $\lambda = 2\pi/k$ is shorter than the ion cyclotron radius $\rho_s$, the energy spectrum is approximately proportional to $E_k \propto \sim k^{4\Delta_\psi - 2\Delta_\phi + 1}$.

In addition, following the idea of the constant enstrophy flux condition [17, 18], the following condition is imposed in the inertial range:

$$\frac{\partial <\Omega>}{\partial t} = constant \quad (10)$$

The constant enstrophy flux condition for the H-M equation in Eq. (2) is rewritten as follows:

$$\frac{\partial <\Omega>}{\partial t} = \iint d^2 x \, \langle \nabla^2 \varphi \frac{\partial}{\partial t}(\varphi - \nabla^2 \varphi)\rangle \quad (11)$$

The local enstrophy flux constant condition becomes as follows:

$$\langle \nabla^2 \varphi \frac{\partial}{\partial t}(\varphi - \nabla^2 \varphi)\rangle = constant \quad (12)$$

We study the steady state described by Eq. (2). When we consider a uniform turbulence in plasma, from the H-M equation Eq. (1), Eq. (12) means the following:

$$\langle \nabla^2 \varphi \frac{\partial}{\partial t}(\varphi - \nabla^2 \varphi)\rangle = \langle \nabla^2 \varphi \{\nabla \varphi \times \hat{z} \cdot \nabla(\nabla^2 \varphi)\}\rangle$$
$$= -\langle(\partial_\lambda \psi^\lambda)\{\psi_\mu \partial_\nu(\partial_\xi \psi^\xi)\}\rangle$$
$$\propto -< L_{-1}\psi \cdot (L_{-1}^2 - \bar{L}_{-1}^2)\phi >$$
$$\propto \delta z^{-2(\Delta_\psi + 1 + \Delta_\phi + 2) + 2\Delta_\chi} < \chi >$$
$$= constant = \delta z^0 \quad (13)$$

Here we include the contribution of the one-point function in our non-unitary solutions, and $\chi$ is the primary field with the dimension of $\Delta_\chi$ coming from the OPE of $[\psi][\phi]$. In Eq. (13), $L_{-n}$ shows the Virasoro generator [13]. Therefore, the constant enstrophy condition becomes as follows for the H-M equation (1):

$$\Delta_\psi + \Delta_\phi + 3 - \Delta_\chi = 0 \quad (14)$$

The condition for the constant enstrophy flux in Eq. (14) is the same one found in Refs. [15, 18], though the primary field is not the electrostatic potential of $\varphi$ but $\psi_\mu \equiv \partial_\mu \varphi$ in this work. In Refs. [15, 18], the primary field was the stream function. This fact may be quite natural, because the H-M equation is based on the fluid model for plasma with a static magnetic field and the static electric field.

Accordingly, the resultant CFT conditions for fluid turbulence in Ref. [18] also fulfil the requirements and conditions found again in this work for the H-M plasma turbulence, except the energy spectrum exponent in Eq. [9]. In Table 1, we show again several example results for the minimal models and the new exponents for each model. The central charge $C$ is also shown by $C = 1 - 6(p - q)^2/pq$ for the minimal model $\mathcal{M}_{p,q}$. In this work in the CFT framework, a number of possible solutions are found under the conditions obtained.

The CFT results found in this work may propose a few possibilities, at least the following two possibilities, to understand the results obtained. One of them is as follows: The results presented in this work suggest that plasma steady turbulence governed by the H-M equation of Eq. (1) may not just have a single exponent but may have multiple exponents for the energy spectrum. The multiple-valued exponents of the energy spectrum may come from the differences in the initial conditions [19] or in boundary conditions. For example, in Ref. [19] the 2D fluid turbulence would represent an interesting dependence of the energy spectrum exponent on the initial conditions. Depending on the initial condition, the energy spectrum exponent would change in the steady state. This characteristic of multiple values for the energy spectrum exponents can be tested by experiments in future.

There would be another possibility for he CFT results in the H-M plasma turbulence and in the fluid turbulence. CFT itself would not perfectly clarify physics involved by each minimal model. As shown above, the work in this paper presented that the CFT conditions and requirements obtained are the same except the energy spectrum exponent for the fluid turbulence and plasma turbulence by the H-M equation. This fact means that CTF results may cover wider physics concerned, and in order to specify one concrete result for a specific physics, we may need additional conditions or consideration or works.

Here we should also point out that it may not be clear if CFT shows exact results for specific plasma turbulence. In this work the dimension analyses are performed in the 2D CFT without the details of the Lagrangian for the H-M plasma turbulence, because the Lagrangian density for the H-M equation is not perfectly clarified yet [20]. In addition, in many works [14, 15, 18, 21, 22] including this work the minimal models are employed to study turbulence. However, 2D CFT would have many other possible models, based on the infinite-dimensional symmetry in CFT. This may also expand the possible solution window for plasma turbulence.

Table 1. Several example results for the Hasegawa-Mima plasma turbulence. The energy spectrum exponent $\alpha = 4\Delta_\psi - 2\Delta_\phi - 1$ shows the exponent of $E_k \propto (1+k^2)k^\alpha$

| $\mathcal{M}(p,q)$ | $\psi$ | $\phi$ | $\chi$ | $C$ | $\alpha$: Energy exponent |
|---|---|---|---|---|---|
| (2,33) | (1,10) | (1,17) | (1,16) | -86.36 | -5.727 |
| (3,43) | (1,12) | (1,15) | (1,14) | -73.42 | -6.837 |
| (4,59) | (2,26) | (1,15) | (2,30) | -75.91 | -6.581 |
| (5,71) | (4,55) | (1,15) | (4,57) | -72.62 | -6.930 |
| (6,55) | (4,34) | (1,11) | (4,44) | -42.65 | -4.045 |

Acknowledgements

The work was partly supported by Utsunomiya University, Japan, Shanghai Jiao Tong University and A program of high-end foreign expert introduction plan, China.